\documentstyle[aps,preprint,epsfig]{revtex}
\tightenlines
\begin{document}
\draft

\title{
New mechanism of X-ray radiation from a relativistic charged particle in
a dielectric random medium }

\author{
Zh.S. Gevorkian$^{1,2}$\footnote{E-mail:~gevorg@stp1.phys.sinica.edu.tw}, C.P. Chen$^{1}$\footnote{Corresponding author, E-mail:~cpchen@phys.sinica.edu.tw}, and Chin-Kun Hu$^{1}$
}
\address{
${1})$ Institute of Physics, Academia Sinica, Nankang, Taipei 11529, Taiwan \\
${2})$ Yerevan Physics Institute and Institute of Radiophysics and Electronics
378410, Ashtarak-2, Armenia\\
}

\date{\today}
\maketitle

\begin{abstract}

We have considered X-ray radiation of a relativistic charged particle moving
in a system consisting of microspheres distributed randomly
in a dielectric material. A new mechanism based on the diffusional
scattering of pseudophotons is suggested. The result leads to a stronger
dependence of radiation intensity on the particle energy, $\gamma$ = E/mc$^2$,
than that predicted by the traditional transition radiation theory,
and explains a recent experiment on such a system of randomly distributed
superconducting granules.
\vskip 0.5cm
{PACS number : 41.60.-m, 61.80.Cb, 85.25.-j}
\\
\end{abstract}

\pagebreak

Radiation of charged particles in randomly inhomogeneous media is a problem
of current interests. One of the applications is to utilize this as a
detector to determine the energy of a relativistic charged particle since
radiation of this nature strongly depends on the particle energy, $\gamma$.
The feature of $\gamma$ dependence makes the detection sensitivity increase
as the particle energy goes up. The currently existing detectors, such as
ionization chamber, Cerenkov detector etc., measure the particle energy
by measuring the velocity, $\beta$ = v/c. This loses the sensitivity since
in the relativistic regime, $\beta$ approaches 1.
\par
One type of the cryogenic detectors is qualified to belong to the category
of random system. It is consisting of type-I superconducting microspheres
randomly distributed in dielectric material such as wax or varnish etc.
In the H-T phase diagram, there is a superheated superconducting region above
the critical transition line. It is a metastable state in which the
superconducting state persists under the condition that no external nucleation
disturbances exist even if it goes across the critical transition
line from the superconducting to the normal state region. The critical field
which confines the superheating state is the superheating field, H$_{sh}$.
The superconducting microsphere in the superheating state will undergo
irreversible phase transition to the normal state if an external energy
exceeding the transition threshold energy is deposited on the grain. By
detecting the grain transition signal, one can actually detect the energy
deposited on the grain. For a detail description on the superconducting
granule device, see, for example, the references listed in \cite{Pretzl}.
\par
Recently, Yuan et al carried out an experiment utilizing the type of detector
as described above to investigate the X-ray radiation from a relativistic
charged particle passing through the detector \cite{Yuan}. The
detector is consisting of superconducting Sn microspheres randomly distributed
in paraffin wax to form a cylindrical bar of 0.3 cm in diameter and 1.2 cm in
length. The microsphere is 35 $\mu$m in diameter. The experimental result shows
that the number of grain transitions, hence the number of X-ray photons
emitted, is proportional to $\gamma$.
\par
By the conventional Transition Radiation (TR) theory \cite{Garibian},
the charged particle
will emit photons when passing across the interface. One can calculate the
radiation intensity or the number spectrum of photons emitted by summing the
contributions from the interfaces. This method of calculation takes each local
interface effect into account as commonly applied in the conventional thin foil
TR detector. According to the TR theory, the intensity emitted is linearly
proportional to $\gamma$, while the number of photons generated,
proportional to ln($\gamma$), see for example, reference \cite{Mikaelian}.
\par
Here, we explain the result of the experiment carried out by Yuan et al. based
on the diffusional scattering of pseudophotons.
The idea is to view the electromagnetic field
associated with a moving charged particle as a collection of pseudophotons.
When the charged particle moves in a medium of inhomogeneous dielectric
property, the pseudophoton can be scattered to become real photon and
detected in an experiment. From this point of
view, the cause of the radiation is not confined to the local interface where
the charged particle passes, as described in the TR theory, but, instead,
resulting from the neighbouring region next to the particle trajectory
with inhomogeneous dielectric constant.
This makes the radiation a global effect rather than a local one.
\par
The number of pseudophotons associated with the moving charged particle
is estimated as follows,

\begin{equation}
N_{ps} \sim \int\frac{d\vec{q}}{(2\pi)^3}A^2(\vec{q})    \label{Nps1}\;\;\; ,
\end{equation}
where $\vec{A}(\vec{q})$ is the vector potential of the electric field
created by the moving charge. The Maxwell's equation for this vector
potential has the form,

\begin{equation}
\label{Max1}
\nabla^2\vec{A} + \frac{\epsilon\omega^2}{c^2}\vec{A} = - \frac{4\pi}{c}\vec{j}(\vec{r},\omega) \;\;\; ,
\end{equation}
where
\begin{equation}
\label{Max2}
\vec{j}(\vec{r},\omega)=\frac{e\vec{v}}{v}\delta(x)\delta(y)e^{i\omega z/ v}
\;\;\; ,
\end{equation}
is the current of a charged particle moving uniformly in the z-direction
with velocity v, and $\epsilon$ is the average dielectric constant of the
system, see below for the details. From the symmetry of the
problem, it follows that the vector potential is directed along the
z-direction, A$_{i}$=A$_{\hat{z}i}$. Then, the Eq. (\ref{Max1}) is easily
solved to have the solution,
\begin{equation}
\label{Max3}
A(\vec{q}) = \frac{-8\pi^2 e}{c}\frac{\delta(q_z - \omega/v)}{k^2 - q^2}\;\;\; ,
\end{equation}
where $k=\omega\sqrt{\epsilon}/c$ is the wavenumber of the pseudopohotons.

By substituting Eq. (\ref{Max3}) into Eq. (\ref{Nps1}), and replacing
$\delta$(0) by $L_z$/2$\pi$ to take into account the finite size of the system,
one obtains

\begin{equation}
N_{ps}(\omega) \sim \frac{e^2}{c}\gamma^2_{m}L_z \label{Nps2}\;\;\; ,
\end{equation}
where $\gamma_m = (1-\beta^2\epsilon)^{-1/2}$ is the Lorentz factor in
the medium in which we believe that v$\sqrt{\epsilon} < $c is always the
case, and L$_z$ is the length of the system.
\par
A simple sketch of the superconducting granule detector is presented in
Fig. 1. The pseudophoton field is moving together with the charged particle
as plotted. Once the charged particle along with the pseudophotons enter the
system, they experience the variation of the dielectric constant introduced
by the grains distributed in the wax and show scattering effect. Since the
pseudophoton is moving along the z direction and in the X-ray region
under consideration, the wave length is much smaller than the grain diameter,
$\lambda << a$, the pseudophoton experiences the randomness in the
moving direction without the effect resulting from the transverse direction.
The randomness of the dielectric constant
as seen by the pseudophoton is therefore of the same nature
for the granular system as for the system of plates of thickness $a$ oriented
perpendicular to the moving direction of the charge. The only difference is
that the plate thickness $a$ is uni-size, while there is a distribution
in the path length of the pseudophoton when traversing the granule. This
introduces further randomness which does not modify the result qualitatively.
So, we can describe the superconducting granule system
by the model system consisting of a homogeneous medium with dielectric
constant $\epsilon_0(\omega)$ in which parallel plates of thickness $a$ and
dielectric constant $\epsilon_p(\omega)$ are randomly spaced in the
z-direction. This is a system with dielectric constant of 1-D randomness
as proposed by Gevorkian \cite{Gevorkian}. When a charged particle
moves in such a system in the z-direction, the surrounding pseudophoton
scatters from the inhomogeneity of the dielectric constant and is converted
into a real photon. Scattering of this kind is characterized by the
scattering mean free path in the z-direction, which is
related to the randomness of the media via the correlation function of the
dielectric constant. Under the condition of Born approximation,
$ka|\sqrt{\epsilon_p/\epsilon_0}-1|$ $<<$ 1, the mean free path is estimated as

\begin{equation}
\label{mfp1}
l(\omega) = \frac{4c^2}{na^2(\epsilon_p-\epsilon_0)^2\omega^2} \;\;\; ,
\end{equation}
where $n$ is the concentration of plates in the z-direction, $\epsilon_p$
and $\epsilon_0$ are the dielectric constants for the plate and the
homogeneous medium respectively.


The multiple scattering occurs under the condition that
$\lambda << l(\lambda) << $L, $l_{in}$($\lambda$). L is the characteristic
dimension of the system and $l_{in}$ is the inelastic mean free path of
pseudophoton. This condition leads to the diffusion of pseudophoton in the
z-direction. The radiation intensity in the diffusion scattering region,
integrated over the angles, is estimated as follows,

\begin{equation}
\label{radiation1}
I(\omega) \sim N_{ps}\frac{1}{l(\omega)}\frac{L^2}{l^2(\omega)}\;\;\;  .
\end{equation}           
It is proportional to the number of pseudophotons, N$_{ps}$, the 
probability of pseudophoton scattering, 1/$l$($\omega$), and the mean number 
of scatterings in the random medium, L$^2$/$l^2$. By substituting 
Eq. (\ref{Nps2}), the radiation intensity in Eq. (\ref{radiation1}) 
becomes,

\begin{equation}
\label{radiation2}
I(\omega) \sim \frac{e^2}{c}\gamma^2_{m}\frac{L_z}{l(\omega)}\frac{L^2}{l^2(\omega)} \;\;\; .
\end{equation}  

In the emission of X-ray, the plasma formula for the dielectric constant 
applies.

\begin{eqnarray}
\label{dielectric1}
\epsilon_p(\omega) & \approx & 1-\frac{\omega^2_p}{\omega^2}\;\;\; ,  \nonumber  \\
\epsilon_0(\omega) & \approx & 1-\frac{\omega^2_{p0}}{\omega^2}\;\;\; ,
\end{eqnarray}
where $\omega_p$ is the plasma frequency of the plate, and $\omega_{p0}$, 
of the homogeneous medium. The average dielectric constant of the system,
$\epsilon$ = $\epsilon_0$ + $na(\epsilon_p-\epsilon_0)$, then becomes,

\begin{equation}
\label{dielectric2}
\epsilon(\omega) \approx 1 - \frac{(1-na)\omega^2_{p0}+(na)\omega^2_p}{\omega^2}\;\;\; ,
\end{equation}
and the Lorentz factor in the medium has the following form accordingly,

\begin{equation}
\label{gamma1}
\gamma_m = \gamma\left(1+\frac{\omega^2_{c}}{\omega^2}\right)^{-1/2} \;\;\; .
\end{equation}
 
In the above equation, one defines the characteristic frequency 

\begin{equation}
\label{frq1}
\omega_{c} = \gamma\sqrt{(na)\omega^2_p + (1-na)\omega^2_{p0}}\;\;\; .
\end{equation}
This characteristic frequency is a result of the overall dielectric 
inhomogeneity and the particle energy, $\gamma$. Substituting the plasma 
formulae, Eq. (\ref{dielectric1}), into the 
condition of Born approximation, one obtains,

\begin{equation}
\label{freq2}
\omega >> \frac{a(\omega^2_p - \omega^2_{p0})}{2c}\equiv \omega_b\;\;\; .
\end{equation}
The elastic mean free path of the pseudophoton, then, takes the form,

\begin{equation}
\label{mfp2}
l(\omega) \approx \frac{1}{n}(\frac{\omega}{\omega_b})^2 \;\;\; . 
\end{equation}
The characteristic frequency, $\omega_b$, is related to the plate thickness 
and the dielectric property across the boundary of the plate surface.
\par
Weak absorption of pseudophotons can be taken into account in similarity 
to the case of wave propagation \cite{Anderson}. The electromagnetic field, 
or the pseudophoton, 
can be absorbed by the random medium. In the weak absorption limit, the 
elastic mean free path is much smaller than the inelastic one, 
$l << l_{in}$($\omega$). Under diffusion propagation of pseudophotons, 
the quantity $\sqrt{l\cdot l_{in}}$, instead of L, becomes the effective 
size of the system provided that $\sqrt{l\cdot l_{in}}$ $\leq$ L,
\cite{Anderson}. Therefore, the radiation intensity becomes,

\begin{equation}
\label{mfp3}
I(\omega) \sim \frac{e^2}{c}\frac{\gamma^2}{(1+\frac{\omega^2_{c}}{\omega^2})}\frac{L_z l_{in}(\omega)}{l^2(\omega)}\;\;\; .
\end{equation}
The ratio, $l_{in}$/$l$, is the mean number of pseudophoton scatterings before 
the absorption. The inelastic mean free path of pseudophoton in the X-ray 
region is determined by the absorption mainly via photoelectric effect, 

\begin{equation}
\label{mfp4}
l^{-1}_{in}(\omega) = (1-na)N_0\sigma^0_{ph}(\omega)+(na)N_p\sigma^p_{ph}(\omega)\;\;\; ,
\end{equation}
where N$_0$ and N$_p$ are the atom numbers in unit volume of the homogeneous 
medium and of the plate respectively. $\sigma^0_{ph}(\omega)$ and 
$\sigma^p_{ph}(\omega)$ are the corresponding photoelectric cross sections. 

The frequency dependence of the radiation intensity in different frequency 
regimes is derived as follows. In the region, $\omega_b << \omega << \omega_L$ 
(where $\omega_L$ is the frequency at which the effective size 
equals the system characteristic size, 
$\sqrt{l(\omega_L)\cdot l_{in}(\omega_L)}$ = L), the absorption of the 
pseudophoton determines the effective size of the system. Therefore, 
taking into account that in this region the photoabsorption cross section is
hydrogen-like, $\sigma_{ph}(\omega) \sim \omega^{-3.5}$, we obtain the 
frequency dependence of radiation intensity, I($\omega$), and the number 
of photon emitted from the detector, N($\omega$)=I($\omega$)/$\hbar\omega$, 
from Eqs. (\ref{mfp3}) and (\ref{mfp4}),   

\begin{equation}
\label{radiation3}
I(\omega) \sim \frac{e^2}{c}\gamma^2\frac{\omega^{-1/2}}{1+\frac{\omega^2_{c}}{\omega^2}}\;\;\; , 
\end{equation}
and
\begin{equation}
\label{N1}
N(\omega) \sim \alpha\gamma^2\frac{\omega^{-3/2}}{1+\frac{\omega^2_{c}}{\omega^2}}\;\;\; .
\end{equation}
For $\omega > \omega_L$, the diffusion trajectory is cut on the system 
characteristic size. We have the formula for I($\omega$) and N($\omega$) 
in this frequency range as,

\begin{equation}
\label{radiation4}
I(\omega) \sim \frac{e^2}{c}\gamma^2\frac{\omega^{-6}}{1+\frac{\omega^2_{c}}{
\omega^2}}\;\;\; , 
\end{equation}
and
\begin{equation}
\label{N2}
N(\omega) \sim \alpha\gamma^2\frac{\omega^{-7}}{1+\frac{\omega^2_{c}}{\omega^
2}}\;\;\; .
\end{equation}
where $\alpha$ = e$^2$/$\hbar$c is the fine structure constant. 

From the above equations, by integrating over frequencies and taking 
into account that $\omega_{c} \propto \gamma$, one obtains the energy 
dependence of the total radiation intensity, I$_{T}$($\gamma$), 
and the total number of emitted photons, N$_{T}$($\gamma$), coming out
of the pseudophoton field in the folllowing two frequency ranges.\\
For $\omega_{c} < \omega_L$,

\begin{eqnarray}
\label{radiation5}
I_T(\gamma) & \sim & \gamma^{2.5} \;\;\; ,\nonumber \\
N_T(\gamma) & \sim & \gamma^{1.5} \;\;\; .
\end{eqnarray}
For $\omega_{c} > \omega_L$, the radiation reaches the saturation region,

\begin{eqnarray}
\label{radiation6}
I_T(\gamma) & \sim & Constant \;\;\; ,\nonumber \\
N_T(\gamma) & \sim & Constant \;\;\; . 
\end{eqnarray}
Since the frequency, $\omega_L$, is a signature which marks that the 
effective size of the system under diffusional propagation of pseudophoton 
equals the system size, the radiation intensity of the charged particle 
is saturated once the pseudophoton field is cut by the system size spatially. 
The saturation region is therefore determined by the system size L$^2$ and 
the particle 
energy, $\gamma$. From $l(\omega_L)\cdot l_{in}(\omega_L) = L^2$, together 
with $l_{in}$($\omega_L$) $\propto \omega^{3.5}_L$ and 
$l(\omega_L) \propto \omega^2_L$, we have $\omega_L \propto$ L$^{4/11}$. 
\par
In formulae (\ref{radiation5}) and (\ref{radiation6}), we have found the energy 
dependence of the emitted radiation intensity and the number of emitted 
photons. On the other hand, in the experiment \cite{Yuan}, the number of 
absorbed photons is measured. These two are related to each other if we 
consider the photon emitted from the pseudophoton field as well as the 
photon absorbed by the system at the same time. Since the effective linear 
size of the system is $\sqrt{l\cdot l_{in}}$, 
there exist many such subsystems in the medium along the path of moving 
electron in the z-direction. After real photons are formed with a spatial 
extent of $\sqrt{l\cdot l_{in}}$, they propagate through the rest of the 
system with part of them absorbed by the system. Under the condition of 
weak absorption, the number of real photons absorbed is proportional to the 
total number of real photons and the absorption probability of a diffusing 
photon determined by $\sqrt{l\cdot l_{in}}$/$l_{in}$ in a system 
of effective size $\sqrt{l\cdot l_{in}}$. Therefore, one can evaluate the 
total number of absorbed photons as,

\begin{equation}
\label{N3}
N_{ab} \leq \int^{\omega_{c}}_{\omega_b} d\omega\left[\frac{l(\omega)}{l_{in}(\omega)}\right]^{1/2} N_T(\omega) \;\;\; .
\end{equation}
By taking into account the frequency dependence of $\omega_{c}$, 
$l$($\omega$), $l_{in}$($\omega$), and N$_T$($\omega$), we obtain from 
Eq. ($\ref{N3}$), N$_{ab} \sim \gamma^x$, with x $\geq$ 3/4. Also,
from Eq. (\ref{radiation5}), the upper limit of the exponent is 1.5. So, 
the number of grain transitions 
has a $\gamma$ dependence to the power within the bound $0.75 \leq x\leq 1.5$.
This puts the experimental result \cite{Yuan} within this bound. 
\par
We now estimate some of the quantities obtained in the above equation for the 
system of the cryogenic detector used in the experiment by 
Yuan et al \cite{Yuan}. The volumetric filling factor of 13 \% for the Sn
grains mixed in the wax corresponds to $n \sim$ 130 cm$^{-1}$. 
By identifying the grain diameter as the plate thickness in the above model, 
$a$ $\sim$ 35$\mu$m, one has $na$ $\sim$ 0.46. The characteristic size of 
the system is roughly 0.3 cm, which is the diameter of the detector cylinder 
in this case. Using $\omega_p \sim$ 50 eV for Sn and $\omega_{p0} \sim$ 20 eV 
for wax, we have $\omega_b 
\sim$ 120 KeV and $\omega_{c} \sim$ (37eV)$\gamma$ which ranges from 
74 to 740 KeV corresponding to the beam energy from 1 to 10 GeV. From
the characteristic frequency above, one can estimate the elastic 
mean free path in the frequency range of interests, which is about a few 
hundred KeV. Indeed, with the frequency below $\omega_L$, which is estimated 
to be a few hundred KeV, the elastic mean free path, $l$($\omega$), is from 
0.02 up to 0.3 cm corresponding to photon energy from 200 to 800 KeV. 
If we calculate $\sqrt{l\cdot l_{in}}$ from the elastic mean free path, $l$, 
obtained above together with the $l_{in}$ calculated from the mass energy 
absorption coefficient \cite{Seltzer}, we obtain the effective size of the
system being from 0.1 cm to the order of 1 cm within this energy range. 
This is about the size of the characteristic dimension of the system. 
So, we have $\sqrt{l\cdot l_{in}} \sim L$. This puts the detector 
system under consideration in the right range for description
by the diffusional emission mechanism, and the result gives a correct trend
on the response of the detector subject to the beam test in terms of 
grain transition. 

In conclusion, we have suggested a new mechanism of X-ray radiation 
from a relativistic charged particle moving in a random medium based 
on the multiple scattering of electromagnetic field (pseudophoton). 
Strong energy dependence of the number of superheated superconducting 
grains proceeding the phase transition to the normal state, observed 
in the experiment \cite{Yuan}, is a 
manifestation of this radiation mechanism. Experiment on the radiation 
from the system of randomly spaced 
plates has been performed by Detoeuf et al \cite{Detoeuf}. However,
the radiation investigated is in the region of soft X-ray from 6 to 12 KeV,
which is out of the region valid for diffusional scattering because 
the absorption is strong so that
$l(\lambda) << l_{in}(\lambda)$ is not satisfied.
For the future experiment to provide more evidences on the diffusional
mechanism of X-ray radiation, we propose criteria for the diffusional region,
$\lambda << l << L, l_{in}$. The X-ray radiation is greatly enhanced 
in this region. A detector working on this would then have a promising 
application in measuring the energy, $\gamma$, of an extremely relativistic 
charged particle.
\par
This work was supported in part by the National Science Council of the Republic 
of China (Taiwan) under contract No. NSC 89-2112-M-001-005.

\begin{figure}[p]
\centerline{\epsfig{file=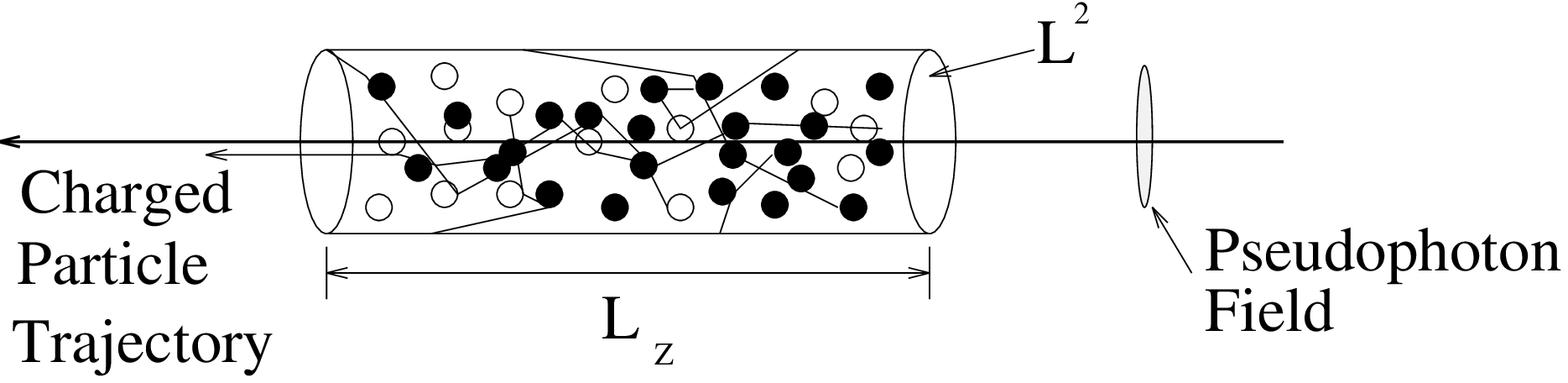,width=18cm,angle=270}}
\caption{Superconducting granule detector. Solid circles represent the Sn
grains in superheating state, while the open ones, for those in normal
state. The grain diameter is $a$. Trajectories for the scattered
pseudophotons are plotted}
\end{figure}

\end{document}